\documentclass[aps,prl,reprint,superscriptaddress,nofootinbib,floatfix,longbibliography]{revtex4-1}
\usepackage[dvipsnames]{xcolor}
\usepackage{amssymb,amsmath,amsfonts,bm,slashed,soul,ragged2e,graphicx,epstopdf,hyperref,array,gensymb}
\usepackage[utf8]{inputenc}
\usepackage[normalem]{ulem}
\usepackage[shortlabels]{enumitem}
\usepackage{colortbl,color,caption,subcaption}
\enlargethispage{\baselineskip}
\definecolor{steelblue}{RGB}{25,25,112}
\definecolor{dullblue}{rgb}{0,0.298,0.49}
\definecolor{darkred}{rgb}{0.545,0,0}
\definecolor{darkorange}{RGB}{222,132,69}
\definecolor{darkgreen}{RGB}{126,171,85}
\definecolor{blue2}{cmyk}{1, 0.1, 0.1, 0}
\hypersetup{colorlinks,linkcolor={darkred},citecolor={blue},urlcolor={dullblue}}
\DeclareCaptionJustification{justified}{\justifying}
\everymath{\displaystyle}

\usepackage{titlesec}

\usepackage{comment}

\newcommand{\beq}{\begin{equation}}
\newcommand{\eeq}{\end{equation}}
\newcommand{\bea}{\begin{eqnarray}}
\newcommand{\eea}{\end{eqnarray}}

\newcommand{\gsim}{\lower.7ex\hbox{$\;\stackrel{\textstyle>}{\sim}\;$}}
\newcommand{\lsim}{\lower.7ex\hbox{$\;\stackrel{\textstyle<}{\sim}\;$}}

\newcommand{\be}{\begin{equation}}
\newcommand{\ee}{\end{equation}}
\newcommand{\ba}{\begin{eqnarray}}
\newcommand{\ea}{\end{eqnarray}}

\def \D {{\rm d}}
\begin{document}

\title{Constraints on Symmetric Dark Matter from Neutron Star Capture and Collapse
}

\author{Yuxin Liu}
\email{liuyuxin211@mails.ucas.ac.cn}
\affiliation{International Center for Theoretical Physics Asia-Pacific, University of Chinese Academy of Sciences, Beijing 100190, China}

\author{Zhen Liu}
\email{zliuphys@umn.edu}
\affiliation{School of Physics and Astronomy, University of Minnesota, Minneapolis, MN 55455, USA}

\author{Maxim Pospelov}
\email{pospelov@umn.edu}
\affiliation{School of Physics and Astronomy, University of Minnesota, Minneapolis, MN 55455, USA}
\affiliation{William I. Fine Theoretical Physics Institute, School of Physics and Astronomy,
University of Minnesota, Minneapolis, MN 55455, USA}

\author{Sanjay Reddy}
\email{sareddy@uw.edu}
\affiliation{
Institute for Nuclear Theory, University of Washington, Seattle, WA 98195, USA}

\begin{abstract}

Dark matter (DM) models with a conserved particle$-$antiparticle number, $n_\chi-n_{\tilde \chi}$, and the asymmetry in the cosmological abundance $n_\chi\neq n_{\tilde \chi}$,  are known to be challenged by the existence of old neutron stars (NSs), as the sufficient accumulation of DM will lead to the collapse of NSs into black holes. We demonstrate that the applicability of these constraints is much wider and covers models with symmetric populations of DM, $n_\chi = n_{\tilde \chi}$, as the process of DM capture regulated by a nucleon-DM scattering can be inherently asymmetric, $\sigma_{\chi n}\neq \sigma_{\tilde\chi n}$. The asymmetry is induced by the interference of different types of $\chi$-$n$ interactions, provided that their combination is odd under charge conjugation in the DM sector, $C_\chi$, and even under combined parity $P_{\chi + n}$. We provide a complete analysis of DM-nucleon bilinear $\chi$-$n$ interactions and find that this asymmetry is very generic. Using canonical NS parameters and local DM halo inputs, we exclude spin‑averaged scattering cross sections down to $\sigma_{n\chi}\!\gtrsim\!10^{-46}\,{\rm cm}^{2}$ at DM mass $m_\chi\!\lesssim\!10^{10}\,{\rm GeV}$ for the maximally asymmetric capture rate, and show that the constraints persist down to very small values of the cross-section asymmetry, ${\cal A}=(\sigma_{\chi n}- \sigma_{\tilde\chi n})/(\sigma_{\chi n}+ \sigma_{\tilde\chi n})\gtrsim 10^{-5}$. 

\end{abstract}

\date{\today}

\maketitle

{\flushleft\underline{\textbf{Introduction:}}} The existence of dark matter (DM) is well established across a wide range of scales, from galactic dynamics to the cosmic microwave background~\cite{Rubin:1970zza,Planck:2018vyg}. A global experimental effort is underway to probe the feeble interactions between DM and ordinary matter. Underground detectors employing tonne-scale liquid noble gas targets now constrain spin-independent DM–nucleon cross sections down to $\sigma_{n\chi} \sim 10^{-47}\,{\rm cm}^2$ for DM masses $m_\chi \sim 100\,{\rm GeV}$~\cite{PandaX-II:2016vec,PandaX-II:2016wea,LUX:2016ggv,LZ:2022lsv,XENON:2023cxc}. However, event rates are suppressed for heavier DM masses as $m_\chi^{-1}$ with fixed local DM energy density. This presents a significant challenge for detecting very heavy DM candidates ($m_\chi \gtrsim 10^5~\text{GeV}$).
Such particles may arise from gravitational production in the early universe~\cite{Chung:1998zb,Chung:1998ua} or from interactions within hidden sectors~\cite{Kramer:2020sbb} that help to evade unitarity bounds~\cite{Griest:1989wd}. Nonetheless, this high-mass regime remains largely unexplored by direct detection experiments, motivating the need for complementary approaches.

Compact astrophysical objects such as neutron stars (NSs) can serve as such a complementary laboratory. Due to their extreme gravitational potentials and nucleon densities $n_{n} \gtrsim 10^{38}~{\rm cm}^{-3}$~\cite{Akmal:1998cf,Haensel:2007yy}, NSs can capture dark matter with cross sections orders of magnitude smaller than those accessible to terrestrial experiments~\cite{Gould:1987ir,Goldman:1989nd,Jungman:1995df,Page:2006ud,Bertoni:2013bsa,Bramante:2023djs,Ema:2024wqr}. Once captured, DM particles thermalize through repeated scattering and accumulate at the stellar core~\cite{Bertoni:2013bsa}, and if DM cannot efficiently annihilate, its self-gravity can eventually overcome quantum and thermal pressure, triggering collapse into a black hole (BH) whose subsequent growth destroys the host NS.

This mechanism has been used to exclude large portions of parameter space in various non-annihilating DM models, based on the observations of old pulsars. Existing literature, see {\em e.g.} \cite{Gould:1989gw,Kouvaris:2007ay,Bertone:2007ae,Bramante:2014zca} identified two necessary conditions that must be fulfilled for the effective accumulation of DM inside NS: 
\begin{enumerate}[$A.$]
\item The DM sector must distinguish between DM ($\chi$) and anti-DM ($\tilde{\chi} $) states, and conserve the number of particles minus antiparticles, $n_\chi-n_{\tilde\chi}$. 
\item An unspecified mechanisms in the early universe must arrange for the \emph{asymmetric} population of DM, $n_\chi-n_{\tilde\chi}\neq0$.
\end{enumerate}
The first condition removes from further considerations ``self-conjugate" dark matter such as real scalars or Majorana fermions. Condition $B$ is far less trivial and requires the dynamical generation of the particle-antiparticle asymmetry in the DM sector, perhaps with some analogy with baryogenesis \cite{Sakharov:1967dj,Huet:1994jb,Ambrosone:2021lsx}. Simplest freeze-in or freeze-out scenarios will not lead to asymmetric DM as $\chi$ and $\tilde \chi$ are created/annihilated in pairs. 

The main observation of this \emph{Letter} is that the class of DM models that are probed by NS$\to$BH transition is in fact much wider and includes particle-antiparticle \emph{symmetric} DM provided that in addition to $A$ we have:
\begin{enumerate}[$B.$]
\item The capture of $\chi$ and $\tilde\chi$ in NS is asymmetric. For the capture based on $\chi-n$ scattering, an asymmetry in the cross section is essential, such that $\sigma_{\chi n} \neq \sigma_{\tilde{\chi} n}$.
\end{enumerate}

In this work, we demonstrate that this condition can be fulfilled quite generically and without very exotic interactions, owing much to the fact that NS is by itself built with $n,p,e$, etc, and not with their antiparticles. At a technical level, the cross section asymmetry, $\sigma_{\chi n} \neq \sigma_{\tilde{\chi} n}$, arises due to interference between $\chi$-$n$ scattering amplitudes with different charge-conjugation and parity properties.  
Specifically, we find that such interference must be odd under both $C_\chi$ (which interchanges $\chi$ and $\tilde \chi$) and $C_\chi P_{\chi + n}$, where $P$ is the common spatial parity transformation on both DM and nucleons. 
 Many examples of similar type asymmetries are known, starting from {\em e.g.} $\sigma_{e^-p}\neq\sigma_{e^+p}$ due to interference of single and double photon exchange \cite{Rachek:2014fam,CLAS:2016fvy,OLYMPUS:2016gso}, (anti)neutrino-nucleon scattering~\cite{Martini:2009uj,Martini:2010ex} and other $CP$-violating phenomena, \cite{Bander:1979px,LHCb:2025ray}.

\begin{figure}
    \centering
    \includegraphics[width=0.45\textwidth]{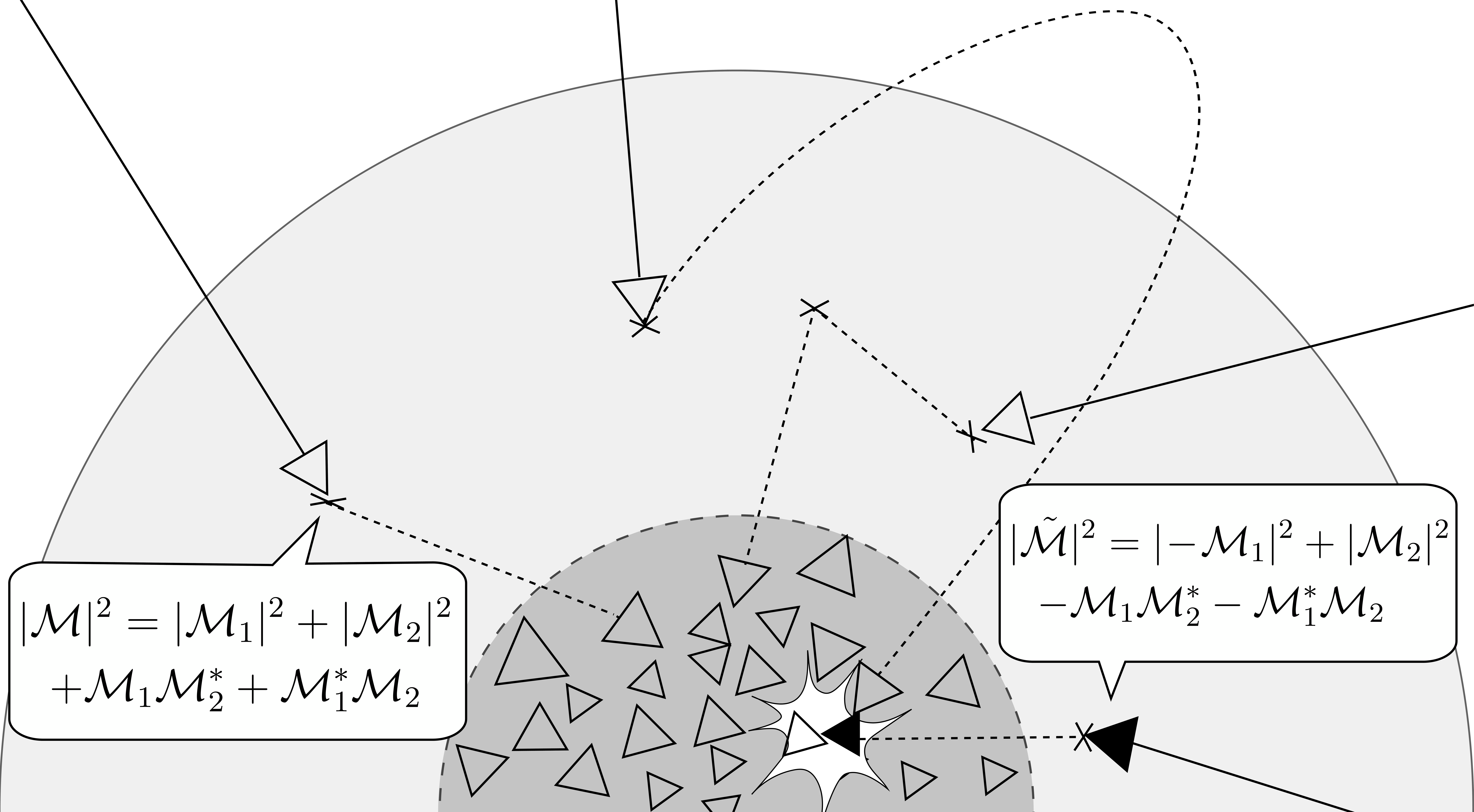}
    \caption{ A schematic diagram illustrating the asymmetric capture of DM and anti-DM by neutron stars. The interference between the two effective operators in the Lagrangian leads to different scattering cross-sections for DM and anti-DM. Asymmetric accumulation of DM in sufficient quantity leads to eventual collapse of NS to BH. }
    \label{fig:schematic}
\end{figure}

This asymmetry in scattering leads to a differential capture rate between $\chi$ and $\tilde \chi$, favoring the accumulation of one species--assumed here to be DM--inside the NS. Over time, this asymmetric capture mechanism builds up a net DM population that may exceed the collapse threshold and induce BH formation, as shown in Fig.~\ref{fig:schematic}. 
While we limit our considerations to relatively short-range interactions,  one could also note that an existence of a long-range $\chi$-$n$ force acting differentially on $\chi$ and $\tilde \chi$ may also result in the differential DM accumulation. 

In the remainder of this \emph{Letter}, we present stringent new NS constraints on symmetric fermionic dark matter. Describing $\chi$-$n$ interactions as operators in effective field theory, we identify the necessary operator pairs (odd under $C_\chi$ and $C_\chi P_{\chi + n}$) leading to asymmetric DM capture. Parametrizing the cross section asymmetry, ${\cal A}$,  we derive constraints on $\{m_\chi,\sigma_{\chi n}, {\cal A}\}$
parameter space from the existence of old NSs with $t_{\rm age}\approx 10^{10}$ yr. The resulting limits apply to $m_\chi \gtrsim 10^7\,{\rm GeV}$, and extend far beyond current and planned DM detection experiments.

{\flushleft\underline{\textbf{Interference-induced Asymmetry:}}}
    
We characterize the interaction between fermionic DM and nucleons through the following effective interactions: 
\begin{equation}
    \mathcal{L}_{\rm int} =\sum_j g_{j}\mathcal{O}^{\chi} \mathcal{O}^n=\sum_j g_{j}(\bar\chi\Gamma \chi)(\bar n\Gamma' n)  
   , \label{eq:general-lag}
\end{equation}
where $\mathcal{O}^{\psi}$ represents the DM (nucleon) bilinear operators constructed from the fermionic field $\psi=\chi \,(n)$ and its Dirac conjugate $\bar{\psi}$. Second equality is a particular realization  of $\mathcal{L}_{\rm int} $ at dimension six level with $\Gamma$ and $\Gamma'$ representing Lorentz structures with appropriately contracted indices. Real parameters $g_{j}$ are the Wilson coefficients. We chose the basis for $\Gamma,\Gamma'$ in such a way that $\mathcal{O}^{\psi}$ are the eigenstates of charge conjugation that can be defined to act separately on the DM and nucleon fields. This way Lorentz components of $\Gamma$'s are also eigenstates of spatial parity. In practice it means that our choice of $\Gamma,\Gamma'$ is achiral: $1,i\gamma_5,\gamma_\mu$ etc. Finally, index $j$ labels all distinct $\Gamma \otimes \Gamma'$ combinations. 

NSs consist of a dense population of unpolarized nucleons, which serves as a target for an incoming, also unpolarized, stream of $\chi$ and $\tilde \chi$. Gravitational acceleration brings the relative $\chi$-nucleon energy to several 100 MeV, so that not only $\chi n$ elastic scattering is possible, but also inelastic processes with $\pi,\gamma$ production. We will concentrate on $2\to2$ elastic scattering as most important, but our treatment can be generalized to inelastic scattering as well. The asymmetric capture will occur provided that for the spin averaged momentum-transfer scattering cross sections, the asymmetry defined as
\begin{equation}
    \mathcal{A} = \frac{\sigma_{\chi n} - \sigma_{\tilde{\chi}n}}{\sigma_{\chi n} + \sigma_{\tilde{\chi}n}},
\end{equation}
is nonzero. Since the cross sections are some linear combinations of $g_i g_j$, our task is to identify which pairings would lead to $\mathcal{A}\neq 0$. It is easy to see that this requires

{\flushleft\emph{$C_\chi$-odd combination}}: two $\chi$ operators that belong to interfering terms must have opposite eigenvalues under the DM charge conjugation, $C_\chi(\mathcal{O}_1^\chi \mathcal{O}_2^\chi)=- \mathcal{O}_1^\chi \mathcal{O}_2^\chi$.

{\flushleft\emph{$C_\chi P_{\chi + n}$-odd combination}}:
Due to absence of defined directions and spin polarizations, the interference term must obey parity conservation, or equivalently being $C_\chi P_{\chi + n}$-odd.\\

The two conditions bear some resemblance to Sakharov's criteria for baryogenesis~\cite{Sakharov1967}, but with key differences: the charge conjugation is applied to the DM part only, as the nucleon target is intrinsically baryon-asymmetric.
One can immediately see that these conditions cannot be satisfied in the lowest Born order by $\mathcal{L}_{\rm int}$ that contains a single effective operator with definite $C_\chi$ property ($\pm$), as square of the amplitude ($\propto g_i^2$) will be even under $C_\chi$. 

\begin{table}[h!]
\centering
\renewcommand{\arraystretch}{1.4}
\resizebox{0.48\textwidth}{!}{
\begin{tabular}{c|cccccccc}
$C_\chi$ / $C_\chi P_{\chi+n}$ 
& $\bar{\chi} \chi$ 
& $i \bar{\chi} \gamma^5 \chi$ 
& $\chi^\dagger_s \chi_s$ 
& $i \chi^\dagger_s \overset{\leftrightarrow}{\partial}_\mu \chi_s$ 
& $\bar{\chi} \gamma_\mu \chi$ 
& $\bar{\chi} \gamma_\mu \gamma^5 \chi$ 
& $\bar{\chi} \sigma_{\mu\nu} \chi$ \\
\hline
$\bar{n} n$ 
& $+/+$ & \textcolor{red}{$+/-$} & $+/+$ &  &  &  &  \\
$i \bar{n} \gamma^5 n$ 
& \textcolor{red}{$+/-$} & $+/+$ & \textcolor{red}{$+/-$} &  &  &  &  \\
$\bar{n} \gamma^\mu n$ 
&  &  &  & \textcolor{brown}{$-/-$} & \textcolor{brown}{$-/-$} & \textcolor{red}{$+/-$} &  \\
$\bar{n} \gamma^\mu \gamma^5 n$ 
&  &  &  & \textcolor{blue}{$-/+$} & \textcolor{blue}{$-/+$} & $+/+$ &  \\
$\bar{n} \sigma^{\mu\nu} n$ 
&  &  &  &  &  &  & \textcolor{brown}{$-/-$}
\end{tabular}}
\caption{$C_\chi$ and $C_\chi P$ eigenvalues for $\mathcal{L}_{\rm int}$ with operator dimension~$\leq 6$. [\textcolor{black}{black} + \textcolor{brown}{brown}] or [\textcolor{blue}{blue} + \textcolor{red}{red}] operator pairs can lead to asymmetric capture. Note that antisymmetric tensor operators can also be contracted using a 4d Levi-Civita tensor and  the corresponding eigenvalues for that case are \textcolor{blue}{$-/+$}.
}\label{tab:CandCPparity}
\end{table}

We list the $C_\chi / C_\chi P_{\chi + n}$ eigenvalues for DM-nucleon bilinear operator combinations $\mathcal{O}^{\chi}\mathcal{O}^n$ in Table \ref{tab:CandCPparity}, where Lorentz indices are assumed to be contracted with the metric tensor. Although our discussion focuses on fermionic DM, the analysis can also apply to bilinears constructed from bosonic DM states, and we include them for completeness under $\chi_s$ notation. As evident from the table, there are 14 operator pairings that produce a nonzero $\mathcal{A}$. When considering only fermionic dark matter, this number decreases to 12. Moreover, some of the interference terms identically vanish upon $\chi$ and $n$ spin averages, cutting the number of possible combinations to 8 for fermionic DM. All these non-vanishing 8 pairs have an overall $CP=+1$.

Since only very heavy fermionic DM can seed BH formation, there is another condition that helps to discriminate further between cases with non-zero $\mathcal{A}$ for fermionic DM. In the scattering process, the momentum transfer, however, is mostly limited by $|\vec{q}|\sim m_n c\ll m_\chi$. In addition, the energy transfer is also suppressed compared to $|\vec{q}|$, $ q^0 \sim O(|\vec{q}|^2m_\chi^{-1}$). Upon the spin average, most $\chi$-$\tilde\chi$ asymmetric combinations require an extra power of DM recoil momenta leading to an additional ${\cal A} \propto m_n/m_\chi,|\vec{q}|/m_\chi< 10^{-7}$
suppression when all $g_i$ are assumed to be of similar size. This leaves only two non-suppressed interfering pairs:
\begin{eqnarray}
\label{SurvivingPairs1}
    {\cal O}^\chi_S{\cal O}^n_S~{\rm and}~ {\cal O}^\chi_V{\cal O}^n_V;~{\cal O}^\chi_A{\cal O}^n_A~{\rm and}~ {\cal O}^\chi_T{\cal O}^n_T
\end{eqnarray}
where $\Gamma,\Gamma'= 1,\gamma_\mu,\gamma_\mu\gamma_5$, $\sigma_{\mu\nu}$ are referred to as $S,V,A,T$.

It is easy to calculate the $\chi$-$\tilde \chi$ cross section asymmetry for both of these cases. 
Lagrangian
\begin{equation}
\mathcal{L}_{\rm int}^{\rm SV} =g_{SS}\bar{\chi}\chi \bar{n}n+g_{VV}\bar{\chi}\gamma_\mu\chi \bar{n}\gamma^\mu n.
\label{SV}
\end{equation} 
can be easily realized in models with spontaneous broken $U(1)$ gauge symmetry and the dark Higgs. Such theory would contain a scalar and a vector mediator particles, both giving rise to \textit{spin-independent} interactions. In the limit of non-relativistic scattering, the asymmetry is given by
\begin{equation}
    \mathcal{A}_{\rm SV} \approx \frac{2 g_{SS}g_{VV}}{g_{SS}^2 + g_{VV}^2}. 
\end{equation}
For a general case without a big hierarchy between the mass scales and couplings between these mediators, it is highly plausible to induce sizable asymmetry for the capture process. 
\textit{Spin-dependent} case in (\ref{SurvivingPairs1}) involves interference of the axial-vector  tensor interactions, both reducing to $\vec{s}_\chi \cdot \vec{s}_n$ structures in the non-relativistic limit. With the Lagrangian:
\begin{equation}
\mathcal{L}_{\rm int}^{\rm AT} =g_{\rm AA}\bar{\chi}\gamma_\mu\gamma^5\chi \bar{n}\gamma^\mu \gamma^5n+\frac{1}{2}g_{\rm TT}(\bar{\chi}\sigma_{\mu\nu}\chi) (\bar{n}\sigma^{\mu\nu} n),
\label{AT}
\end{equation} 
the resulting asymmetry for non-relativistic scattering is given by:
\begin{equation}
    \mathcal{A}_{\rm AT} \approx \frac{2 g_{AA} g_{TT}}{g_{AA}^2 +  g_{TT}^2 } . 
\end{equation}
The tensor interaction bears a strong resemblance with interaction of two magnetic dipoles while axial interaction is similar to a massless pseudoscalar exchange. 

Finally, in exact analogy with the electrodynamics example, $\sigma_{e^-p}\neq \sigma_{e^+p}$, if one goes beyond the leading Born approximation, even a single operator $\cal O$ can induce an asymmetry, provided that it satisfies already established criteria. In particular, $\bar{\chi}\gamma_\mu\chi \bar{n}\gamma^\mu n$ and $\bar{\chi}\sigma_{\mu\nu}\chi \bar{n}\sigma_{\mu\nu}n$ are both $C_\chi$-odd, $P$-even.  Each of these operators {\em by itself} will induce a non-zero ${\cal A}$ due to the interference of the first and second order scattering amplitudes. This means that asymmetric capture can originate even from a single mediator. The resulting asymmetry relies on the details of the realization, but generically is loop and momentum ratio suppressed. In the Supplemental Material, we show additional calculation details, including extension of ${\cal L}_{\rm int}$ to higher-dimensional operators as well as exact expressions for asymmetries ${\cal A} $ 
generated by Eqs. (\ref{SV}) and (\ref{AT}).

{\flushleft\underline{\textbf{Dark Matter Capture and Black Hole Formation:}}}

The process by which astrophysical bodies capture dark matter has been extensively studied in the literature~\cite{Gould:1987ir,Goldman:1989nd,Jungman:1995df,Page:2006ud,Bertoni:2013bsa,Garani:2017jcj,Bramante:2017xlb,Acevedo:2020gro,Bramante:2023djs,Bhattacharya:2023stq,Ge:2024prt,Ema:2024wqr}. We concentrate on the fermionic DM, as the phenomenology of bosonic-DM-induced NS$\to$BH transmutation is far more intricate ~\cite{Kouvaris:2011fi,Bramante:2013hn,Bramante:2023djs}, and is left for  future work. 
We follow standard treatment of DM capture, and adapt it to symmetric DM with non-zero ${\cal A}$, assumed to be positive.

For a single, isotropic in the $\chi n$ center of mass frame scattering event, the average energy loss experienced by a DM particle is given by $m_n (1+z)^2 v_{\rm esc}^2$~\cite{Dasgupta:2020dik}, where the DM velocity inside the neutron star is approximated by the escape velocity at the surface of the NS, $v_{\rm esc} = \sqrt{2GM_{*}/R_*} \approx \mathcal{O}(0.6)$, and $1+z = (1 - 2GM_{*}/R_*)^{-1/2}$ is the corresponding gravitational redshift factor. Here $M_*$ and $R_*$ denote the mass and radius of the NS, respectively. Noticeable deviation from the above approximation can occur~\cite{Dasgupta:2020dik,Bramante:2023djs} for mostly forward scattering due to {\em e.g.} a light force mediator.

To be gravitationally captured by a NS, even a small loss of kinetic energy by a $\chi$-particle can prevent its escape to asymptotic infinity, provided that 
\begin{equation}
    \Delta E_\chi> E_{\rm kin}^\infty = m_\chi v_{\rm rel}^2/2 \sim  10^{-6} m_\chi,
\end{equation} 
where $v_{\rm rel}\sim 10^{-3}c$ is the $\chi$-NS relative velocity at large separation. For DM masses $m_\chi \lesssim  10^6 m_n$, this energy can be easily dissipated through a single scattering event. For heavier DM, an average of $E_{\rm kin}^\infty/\Delta E_\chi \sim 10^6 m_n/m_\chi$ scattering events per each trapped $\chi$ particle is required. 
The capture probability, for an effectively thin target regime, can then be modeled via a Poisson distribution:
\begin{equation}
    P_{\rm cap}^\chi \approx \frac{\sigma_{\chi n} }{\sigma_0} \times \min\left\{1, \frac{ E_{\rm kin}^\infty }{\Delta E_\chi} \right\},
\end{equation}
where $\sigma_{n\chi}$ is the DM-nucleon scattering cross-section, and $\sigma_0 \approx \pi R_*^2 m_n/M_* \approx 2.5 \times 10^{-45}~{\rm cm}^2$ is an estimate of the NS's geometric capture cross-section.  Further refinement of these estimates can come from accounting for DM self-interaction~\cite{McDermott:2011jp,Bramante:2013hn}, consideration of velocity distribution~\cite{Carr:1997cn,Bramante:2017xlb} and other effects~\cite{Gould:1987ir,deLavallaz:2010wp,Bertoni:2013bsa}. 

The same treatment applies for the anti-DM capture probability, $P_{\rm cap}^{\tilde{\chi}}$. The total capture rate per unit time for DM and anti-DM in the NS is then:
\begin{equation}
    C_{\chi/\tilde{\chi}} = n_{\chi/\tilde{\chi}} v_{\rm rel}\times \pi b^2 \times P_{\rm cap}^{\chi/\tilde{\chi}},\label{eq:capture_rate}
\end{equation}
and for numerical results we use $v_{\rm rel} \approx 350~{\rm km/s}$ and $n_{\chi/\tilde{\chi}} = n_{\rm DM}/2 \approx m_\chi^{-1}\times 0.2\,{\rm GeV/cm}^3$~\cite{Carr:1997cn,Zhou:2022lar} as a typical input DM number density. The gravitational focusing results in an impact parameter much larger than the NS radius, $b = R_*(1+z)(v_{\rm esc}/v_{\rm rel})\sim 10^3 R_*$.
This analysis equally applies to spin-independent and spin-dependent $\chi$-$n$ scattering.

Once a net population of captured DM particles builds up, we can safely assume that any subsequently captured anti-DM particles will quickly annihilate with resident DM. Therefore, the growth rate of the net number of DM particles is given by the difference between the DM and anti-DM capture rates as described in Eq.~(\ref{eq:capture_rate}):
\begin{equation}
    \frac{\D N_\chi}{\D t} = C_{\chi} - C_{\tilde{\chi}} \overset{P_{\rm cap}^\chi \ll 1}{\longrightarrow} n_{\rm DM} v_{\rm rel}\times  \pi b^2 \times P_{\rm cap}^{\chi} \frac{\mathcal{A}}{1+\mathcal{A}},
\end{equation}
where the second limit is valid in the optically thin regime.
 In the large cross-section regime, where the approximated expression of $P^\chi_{\rm cap}\gsim 1$, the capture probability saturates to unity for both DM and anti-DM, and the asymmetry of cross sections does not result in the preferential DM built-up inside NS.

The captured dark matter particles thermalize within the NS core quickly within $t_{\rm th}$~\cite{Bertoni:2013bsa}, forming a gravitationally bound halo. Once the total mass of captured DM exceeds a critical threshold $M_{\rm crit}$, this halo undergoes catastrophic collapse to form a BH. If the accretion rate of this black hole exceeds its Hawking evaporation rate, the black hole will inevitably grow and quickly consume the entire NS, within a time interval $t_{\rm grow}$.

For the collapse to occur, several conditions must be satisfied: the dark matter must trigger a self-gravitating instability, overcome its own thermal pressure, and reach the Chandrasekhar limit. 
The Chandrasekhar limit sets the critical mass $M_{\rm crit}$ for the dark matter parameter space of interest:
\begin{equation}
    M_{\rm crit} \approx \frac{M_{\rm pl}^3}{m_\chi^2} \approx 1.6 \times 10^{-16} M_\odot \left( \frac{10^8~{\rm GeV}}{m_\chi} \right)^{2},
\end{equation}
where $M_{\rm pl}=G^{-1/2}$ is the Planck mass, and $M_\odot$ is the solar mass. Further review and explicit expressions for the rates of capture, thermalization and collapse can be found in the Supplemental Material. 

{\flushleft\underline{\textbf{Constraints on Symmetric Dark Matter:}}}
Thousands of pulsars have been observed in the Solar System neighborhood~\cite{Manchester:2004bp}. 
Meanwhile, old NSs with ages $t_{\rm age} \approx 10^{10}~{\rm yr}$ are abundant in our Galaxy; for instance, J1719-1438 is located 336~pc away with an estimated age of 11.4~Gyr, and J1737-0811 is located 206~pc away and is believed to be 8.34~Gyr old~\cite{Manchester:2004bp}.

\par Based on these observations, we adopt a typical NS mass $M_* = 1.5M_\odot$, radius $R_* = 12~{\rm km}$, and age $t_{\rm age} = 10^{10}~{\rm yr}$ for our analysis. The survival of such NSs imposes the condition:
\begin{equation}
    \frac{M_{\rm crit}}{m_\chi \, {\rm d}N_\chi/{\rm d}t} + t_{\rm th} + t_{\rm grow} \geq t_{\rm age},
\end{equation}
where the first term represents the accretion time and dominates over the other two timescales in the solar vicinity for the DM parameter space considered. We evaluate this constraint on the $\{m_\chi,\sigma_{\chi n}, {\cal A}\}$ parameter space for the fermionic DM.

\begin{figure}[thbp]
    \centering
    \includegraphics[width=0.495\textwidth]{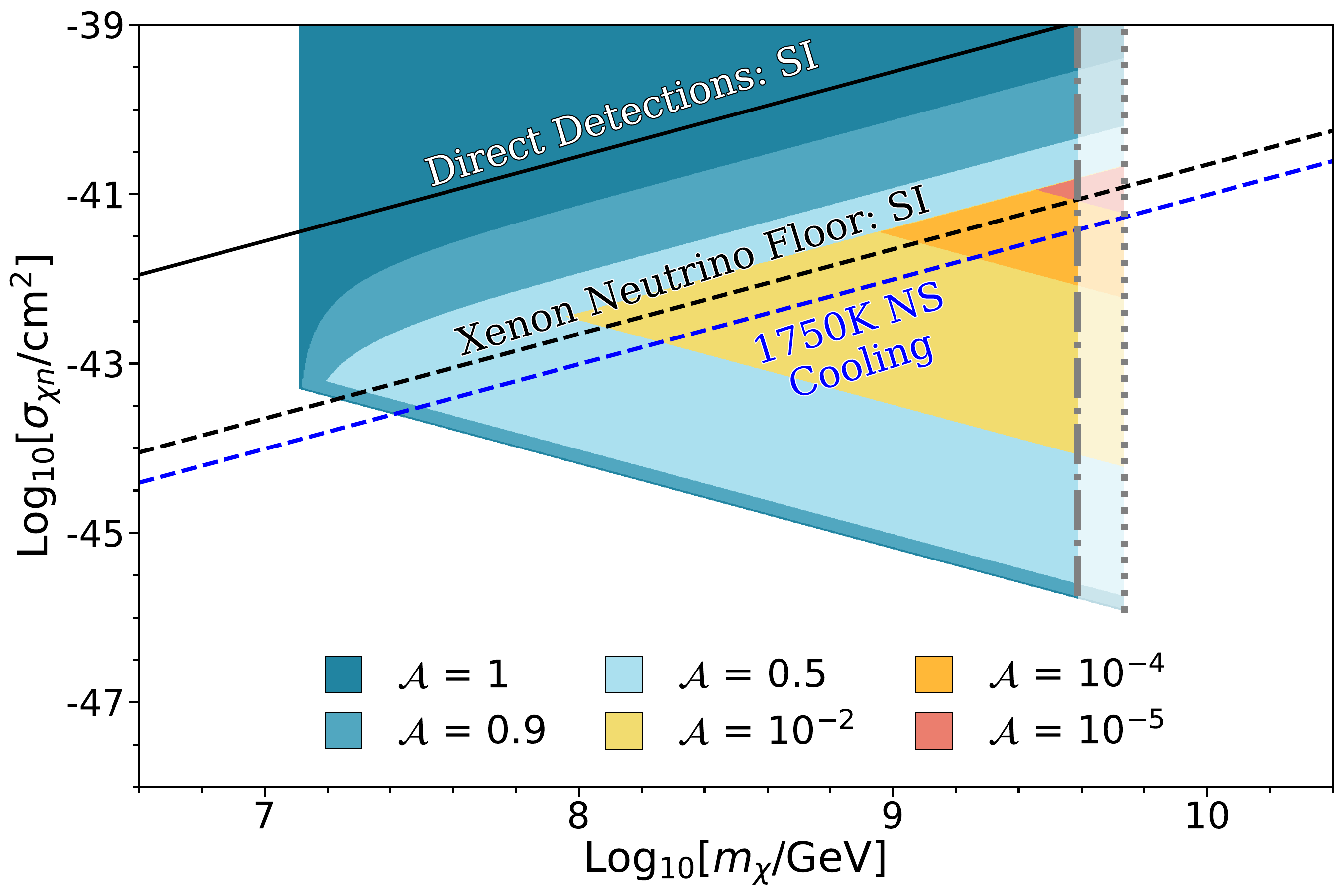}
    \caption{The excluded $\{m_\chi,\sigma_{\chi n}\}$ parameter space for the symmetric fermionic DM model with different input values of the cross section asymmetry $\mathcal{A}$. 
    The vertical gray dotted line and dot-dashed line correspond to two different treatments of the BH evaporation Page factor~\cite{MacGibbon:1991tj,Ray:2023auh}.
    The black solid line represents the current existing constraints from direct detection experiments for spin-independent interactions~\cite{PandaX-II:2016vec,PandaX-II:2016wea,LUX:2016ggv,LZ:2022lsv,XENON:2023cxc}, and the black dashed line indicates irreducible neutrino background (so-called ``neutrino floor" or ``fog") for xenon detectors~\cite{Ruppin:2014bra,OHare:2021utq}. The blue dashed line is a hypothetical NS cooling constraint, assuming an observation of a NS with surface temperature of 1750K and full DM energy-to-heat conversion \cite{Baryakhtar:2017dbj}. 
    }
    \label{fig:exclusion}
\end{figure}

The excluded parameter regions are shown as colored regions in Fig.~\ref{fig:exclusion} for various asymmetry factors $\mathcal{A}$. As expected, the exclusion region shrinks as $\mathcal{A}$ decreases, since a larger fraction of captured DM is annihilated by anti-DM. Moreover, one can recover existing constraints on asymmetric DM scenarios ~\cite{deLavallaz:2010wp} by setting $\mathcal{A} = 1$ in our framework and taking twice the (anti-)DM density used in our calculation.

The sensitivity covers enclosed regions on DM mass vs cross-section plane. To the left and below the shaded region ({\em i.e} for light $m_\chi$ and/or smaller $\sigma_{\chi n}$), the amount of captured DM in the NS lifetime is insufficient to induce gravitational collapse. As the asymmetry factor $\mathcal{A}$ decreases, less net DM can be accumulated, moving the lower-left boundary to the upper right. In contrast, the vertical boundary on the heavier mass side, largely insensitive to $\mathcal{A}$, is due to Hawking evaporation dominating the accretion rate. Such a boundary depends on the BH radiation, and we show the uncertainty due to baseline assumptions by a gray-shaded region. The left boundary (vertical gray dot-dashed line) of the shaded region indicates the consideration of the Hawking radiation rate assuming full SM particle spectrum emission, while the right boundary (vertical gray dotted line) corresponds to only photons and neutrinos being emitted~\cite{MacGibbon:1991tj,Ray:2023auh}. Finally, when the DM-nucleon cross section is big enough, the capture probabilities of both DMs and anti-DMs are nearly $100\%$, making net DM number accumulation impossible.

For a sizable asymmetry ${\cal A}$, the NS limits on $\sigma_{\chi n}$ are generally stronger than constraints from direct detection experiments assuming spin-independent interactions~\cite{PandaX-II:2016vec,PandaX-II:2016wea,LUX:2016ggv,LZ:2022lsv,XENON:2023cxc}, reaching beyond the solar-neutrino-induced direct detection ``fog'' reference value~\cite{Ruppin:2014bra,OHare:2021utq}. Therefore, the NS constraints complement the reach of terrestrial direct detection experiments, particularly at higher DM masses. Improved observations of NSs near/at the Galactic Center, where elevated values of $n_\chi$ are expected, will bring new sensitivity to $\sigma_{\chi n}$ and ${\cal A}$. If ${\cal A} < 10^{-6}$ sensitivity is reached, this would probe a wider spectrum of models compared to the two benchmark cases considered here.

We also show a futuristic benchmark sensitivity from the heating of a NS by capture and annihilation of DM. This mechanism would require an efficient capture of both DM and anti-DM species, corresponding to scenarios with smaller $\mathcal{A}$. Suffice it to say that such a constraint would require a discovery of a NS with  $\lesssim 10^{-5}$ smaller thermal luminosity compared to the coolest NS ever observed~\cite{Guillot:2019ugf}.

{\flushleft\underline{\textbf{Summary and Outlook:}}}
We have demonstrated that the mere existence of old NSs imposes powerful constraints even on \emph{symmetric} DM. This new insight stems from the observation that operator interference leads to an asymmetric capture rate between DM and anti-DM. Our effective field theory framework identifies the minimal $C_\chi$- and $C_\chi P_{\chi+n}$-odd interference amplitudes required to generate such an asymmetry, and we specify ultraviolet completions that realize these features. By applying established results for capture rates, thermalization, and collapse dynamics, we show that the continued existence of NSs in the solar neighborhood already excludes significant regions of parameter space for symmetric DM, complementing terrestrial direct detection experiments.

Our work highlights a gap in the current literature: existing NS constraints have primarily focused on intrinsically asymmetric DM via NS collapse, which requires an additional process in the early universe that would ensure the dominance of DM or anti-DM. 
By showing that interference-driven capture asymmetries open a viable collapse channel even for \emph{symmetric} DM, we extend the ability of NS probes into a new regime of DM models. Our work reinforces neutron stars as unique astrophysical laboratories, whose extreme densities and gravitational fields enable novel detection pathways for DM, with promising prospects for future observational accomplishments. Future radio pulsar surveys, gravitational wave channels, and infrared searches for ultra-cold NSs will strengthen our understanding of NS structures and properties~\cite{Carilli:2004nx,Gardner:2006ky,Nan:2011um,Miller:2023qph}.


\begin{acknowledgments}
\textbf{Acknowledgments} We thank Asher Berlin, Joseph Bramante, Jing Shu, Zhengkang Zhang, and Zizheng Zhou for helpful discussions. The work of ZL and MP is supported in part by the DOE grant \#DESC0011842. ZL is supported in part by a Sloan Research Fellowship from the Alfred P. Sloan Foundation at the University of Minnesota. The work of S.R. was supported by the U.S. DOE under Grant No. DE-FG02-00ER41132.
\end{acknowledgments}

%

\clearpage                      
\appendix                       



\pagebreak
\widetext
\begin{center}
\textbf{\large Supplemental Material:\\Constraints on Symmetric Dark Matter from Neutron Star Capture and Collapse}
\end{center}
\setcounter{equation}{0}
\setcounter{figure}{0}
\setcounter{table}{0}
\setcounter{section}{0}
\makeatletter
\renewcommand{\theequation}{S\arabic{equation}}
\renewcommand{\thesection}{S\arabic{section}}
\renewcommand{\thefigure}{S\arabic{figure}}
\renewcommand{\bibnumfmt}[1]{[#1]}
\renewcommand{\citenumfont}[1]{#1}

\hspace{5mm}

The Supplemental Material provides a more detailed discussion of $C_\chi/C_\chi P_{\chi+n}$ parities of different effective operators and expressions for the asymmetric scattering cross sections in representative scenarios. We also include quantitative descriptions of processes of dark matter (DM) capture, thermalization, collapse to a seed black hole (BH), and its subsequent growth. Throughout this study, we employ natural units $\hbar = c = k_B=1$.

\section{Asymmetric dark matter scattering due to interference}
\par We discuss the asymmetry caused by interference from a most general lagrangian coupling between the DM and the nucleon, including two different low-energy effective current interactions:

\begin{equation}
    \mathcal{L}_{\rm int} \propto g_{1} \mathcal{O}_{1}^{\chi(\mu)}\mathcal{O}_{1(\mu)}^n+g_{2} \mathcal{O}_{2}^{\chi(\nu)}\mathcal{O}_{2(\nu)}^n, \label{eq:general-lag}
\end{equation}
where $\mathcal{O}_{i}^{\chi(\mu)}$ and $\mathcal{O}_{i(\mu)}^n$ are the DM and nucleon bilinear operators, respectively.  We consider elastic scattering process $(\chi/\tilde{\chi}+n\rightarrow \chi/\tilde{\chi}+n)$ with $k$ and $k'$ the initial and final nucleon momentum, and $p$ and $p'$ the initial and final DM (anti-DM) momentum. Mandelstam variables are induced as $s = (p+k)^2, t = (p-p')^2\equiv q^2, u = (p-k')^2$. 
As TABLE.I in the main text shows, the 14 operator pairs that satisfy the $C_\chi$- and $C_\chi P_{\chi+n}$-parity conditions are:
\begin{align}
(1)\quad& {\cal O}^\chi_P{\cal O}^n_P\otimes {\cal O}^\chi_V{\cal O}^n_V;~{\cal O}^\chi_V{\cal O}^n_A\otimes {\cal O}^\chi_P{\cal O}^n_S;~{\cal O}^\chi_V{\cal O}^n_A\otimes {\cal O}^\chi_S{\cal O}^n_P;~{\cal O}^\chi_T\epsilon{\cal O}^n_T\otimes {\cal O}^\chi_A{\cal O}^n_V;~{\cal O}^{\chi_s}_S{\cal O}^n_P\otimes {\cal O}^{\chi_s}_V{\cal O}^n_A,\\
(2)\quad& {\cal O}^\chi_S{\cal O}^n_S\otimes {\cal O}^\chi_T{\cal O}^n_T;~{\cal O}^\chi_P{\cal O}^n_P\otimes {\cal O}^\chi_T{\cal O}^n_T;~{\cal O}^\chi_A{\cal O}^n_A\otimes {\cal O}^\chi_V{\cal O}^n_V;\notag\\
& {\cal O}^\chi_V{\cal O}^n_A\otimes {\cal O}^\chi_A{\cal O}^n_V;~{\cal O}^\chi_T\epsilon{\cal O}^n_T\otimes {\cal O}^\chi_P{\cal O}^n_S;~{\cal O}^\chi_T\epsilon{\cal O}^n_T\otimes {\cal O}^\chi_S{\cal O}^n_P,\\
(3)\quad& {\cal O}^\chi_S{\cal O}^n_S\otimes {\cal O}^\chi_V{\cal O}^n_V;~{\cal O}^\chi_A{\cal O}^n_A\otimes {\cal O}^\chi_T{\cal O}^n_T;~{\cal O}^{\chi_s}_S{\cal O}^n_S\otimes {\cal O}^{\chi_s}_V{\cal O}^n_V, 
\end{align}
where $S,P,V,A,T$ denote scalar $(1)$, pseudoscalar $(i\gamma^5)$, vector$(\gamma^\mu)$, axial-vector$(\gamma^\mu\gamma^5)$, tensor $(\sigma_{\mu\nu})$ bilinears, and $\epsilon$ represents Levi-Civita tensor used to contract two $(\sigma_{\mu\nu})$. However not all of these operator pairs would generate a sizable asymmetric scattering cross-sections, as we will discuss below.

The operators in group (1) vanish exactly, despite the fact that they satisfy the parity conditions. The ${\cal O}^\chi_P{\cal O}^n_P\otimes {\cal O}^\chi_V{\cal O}^n_V$ and $~{\cal O}^\chi_V{\cal O}^n_A\otimes {\cal O}^\chi_P{\cal O}^n_S$ pairs have interference terms which would disappear under the spin-averaging of unpolarized DM and nucleon states. More specifically, on either the DM or nucleon side, the $\gamma_5$ matrix in the pseudoscalar and axial current bilinear would require at least two $\gamma^\mu$ or another $\gamma_5$ matrix in two bilinears together, to avoid vanishing trace. Meanwhile, ${\cal O}^\chi_V{\cal O}^n_A\otimes {\cal O}^\chi_S{\cal O}^n_P$,${\cal O}^\chi_T\epsilon{\cal O}^n_T\otimes {\cal O}^\chi_A{\cal O}^n_V$ and ${\cal O}^{\chi_s}_S{\cal O}^n_P\otimes {\cal O}^{\chi_s}_V{\cal O}^n_A$ have interference terms proportional to $(p+p^\prime)^\mu \cdot(k-k^\prime)_\mu$, which reduces to zero in a $2\rightarrow2$ scattering process under four-momentum conservation.

The operator pairs in group (2) are able to generate asymmetric scattering cross-sections, but the asymmetry is suppressed by the nucleon-DM mass ratio. As explained in the main text, in the Neutron Star (NS) frame, both the DM and nucleon are mildly-relativistic, thus the transferred four-momentum $q_\mu = p^\prime_\mu-p_\mu$ would be limited. Specifically, the recoil energy of DM is at most $\sim m_n^2/m_\chi$, with $|\vec{q}| \lesssim m_n$. If the leading momentum contribution to the interference terms from the DM side is $q^\mu (p_\mu+p^\prime_\mu)$, reducing to $\sim m_\chi m_n$, since all DM operators except $\mathcal{O}^\chi_P$ give $\sim m_\chi^2$ contribution to the symmetric amplitude squared, the asymmetry factor would be suppressed by $\sim m_n/m_\chi$ in the non-relativistic limit, as is the case for all pairs in group (2). Furthermore, the asymmetric cross-section between neutrinos and anti-neutrinos interacting with nucleons is a well-known example of asymmetry caused by interference between axial-vector interactions ~\cite{Martini:2009uj,Martini:2010ex}. However, such asymmetry from chiral coupling ($V\pm A$) is suppressed to DM-nucleon interactions here, as the chiral coupling will induce the suppression $\sim m_n/m_\chi$ as described above. Note that the neutrino case is special, as one does not perform a neutrino spin-average of the cross sections for such a relativistic chiral state. For massive DM scattering, the spin average is performed separately for $\chi$ and $\bar\chi$.

The operators in group (3) would generate an asymmetric scattering cross-sections with $\mathcal{O}(1)$ asymmetry factor $\mathcal{A}$, since the leading contribution to both the symmetric/asymmetric part of amplitude squared and is $\sim m_\chi^2 m_n^2$ for all pairs. We provide the detailed results for both cases below.

{\flushleft{\textbf{Spin-Independent Asymmetric Scattering}}}
For the interaction Lagrangian:
\begin{equation}
    \mathcal{L}_{\rm int}^{\rm SV} =g_{SS}\bar{\chi}\chi \bar{n}n+g_{VV}\bar{\chi}\gamma_\mu\chi \bar{n}\gamma^\mu n,
\end{equation}
one can write down the amplitudes for both DM and anti-DM scattering with nucleons, resulting in the following amplitude squared for DM-nucleon scattering:

\begin{equation}
    \begin{aligned}
        \frac{1}{4}\sum|\mathcal{M}|^2 = &2g_{VV}^2 \left[2m_n^4+2m_n^2(2m_\chi^2-s+t-u)+2m_\chi^4-2m_\chi^2(s-t+u)+s^2+u^2\right]\\
        +&g_{SS}^2\left[(t-4m_n^2)(t-4m_\chi^2)\right]\\
        +&8g_{SS}g_{VV}\left[m_\chi m_n(s-u)\right],
    \end{aligned}\label{eq:ampDM-SS}
\end{equation}

and amplitude squared for anti-DM-nucleon scattering:
\begin{equation}
    \begin{aligned}
        \frac{1}{4}\sum|\tilde{\mathcal{M}}|^2 = &2g_{VV}^2 \left[2m_n^4+2m_n^2(2m_\chi^2-s+t-u)+2m_\chi^4-2m_\chi^2(s-t+u)+s^2+u^2\right]\\
        +&g_{SS}^2\left[(t-4m_n^2)(t-4m_\chi^2)\right]\\
        -&8g_{SS}g_{VV}\left[m_\chi m_n(s-u)\right].
    \end{aligned}
\end{equation}
The $1/4$ factor corresponds to the spin-average of the initial spins. The asymmetric factor $\mathcal{A}$ can be easily calculated. Furthermore, under the non-relativistic limit $q^\mu = p^\mu - p'^\mu < k^\mu\ll p^\mu$, or $s = (m_\chi^2+m_n)^2+|\vec{p}|^2(m_\chi/m_n+m_n/m_\chi)+2|\vec{p}|^2$, $t = -2|\vec{p}|^2(1-\cos{\theta})
$,$u = 2m_\chi^2+2m_n^2-s-t$, with $|\vec{p}|<m_n^2\ll m_\chi^2$ the center of mass momentum magnitude of all the particles, and $\theta$ the scattering angle between the initial DM and nucleon momentum, the asymmetry factor can be simplified as:
\begin{equation}
    \mathcal{A}_{\rm SV} = \frac{\frac{1}{4}\sum|\mathcal{M}|^2 - \frac{1}{4}\sum|\tilde{\mathcal{M}}|^2}{\frac{1}{4}\sum|\mathcal{M}|^2 + \frac{1}{4}\sum|\tilde{\mathcal{M}}|^2} \approx \frac{32 g_{SS}g_{VV} m_\chi^2(2m_n^2+|\vec{p}|^2)}{16g_{VV}^2m_\chi^2(2m_n^2+|\vec{p}|^2(\cos{\theta}+1))+16g_{SS}^2m_\chi^2(2m_n^2+|\vec{p}|^2(1-\cos{\theta}))}\approx \frac{2g_{SS} g_{VV}}{g_{SS}^2+2g_{VV}^2}.
\end{equation}
First approximation takes a non-relativistic limit up to $(v/c)^2$ terms separately for numerator and denominator, and in the second approximation $|\vec{p}|^2 \rightarrow 0$.

{\flushleft{\textbf{Spin-dependent DM-nucleon interaction}}}
Consider a gauge boson mediating between the axial currents of the DM and the nucleon, and another between the tensorial currents of the DM and the nucleon, both integrated out:
\begin{equation}
    \mathcal{L}^{\rm AT}_{\rm int} = g_{AA}\bar{\chi}\gamma_\mu\gamma_5\chi \bar{n}\gamma^\mu\gamma_5 n + \frac{1}{2}g_{TT}(\bar{\chi}\sigma_{\mu\nu}\chi)  (\bar{n}\sigma^{\mu\nu} n),
\end{equation}
where $g_{AA}$ and $g_{TT}$ are the coupling strength of the axial current interaction and the tensorial current interaction, respectively. The non-relativistic operator structure of the two interaction is the same $\sim \vec{s}_\chi\cdot\vec{s}_n$, thus the interference term  survives under spin-averaging. 

Now we also give the fully relativistic amplitude squared for the two interactions:
\begin{equation}
    \begin{aligned}
        \frac{1}{4}\sum|\mathcal{M}|^2 = &2g_{TT}^2 \left[4m_n^4+2m_n^2(8m_\chi^2-4s-t)+4m_\chi^4-2m_\chi^2(4s+t)+(2s+t)^2\right]\\
        +&2g_{AA}^2\left[2m_n^4+4m_n^2(5m_\chi^2-s-t)+2m_\chi^4-4m_\chi^2(s+t)+2s^2+2st+t^2\right]\\
        -&24g_{TT}g_{AA}\left[m_\chi m_n (s-u)\right],
    \end{aligned}
\end{equation}
and amplitude squared for anti-DM-nucleon scattering:
\begin{equation}
    \begin{aligned}
        \frac{1}{4}\sum|\tilde{\mathcal{M}}|^2 = &2g_{TT}^2 \left[4m_n^4+2m_n^2(8m_\chi^2-4s-t)+4m_\chi^4-2m_\chi^2(4s+t)+(2s+t)^2\right]\\
        +&2g_{AA}^2\left[2m_n^4+4m_n^2(5m_\chi^2-s-t)+2m_\chi^4-4m_\chi^2(s+t)+2s^2+2st+t^2\right]\\
        +&24g_{TT}g_{AA}\left[m_\chi m_n (s-u)\right],
    \end{aligned}
\end{equation}
In the same approximations as before, the non-relativistic limit of the asymmetry factor can be calculated as:
\begin{equation}
    \mathcal{A}_{\rm AT} \approx \frac{96g_{TT}g_{AA} m_\chi^2(2m_n^2+|\vec{p}|^2 )}{16g_{TT}^2m_\chi^2(6m_n^2+|\vec{p}|^2(3+\cos{\theta}))+8g_{AA}^2m_\chi^2(12m_n^2+2|\vec{p}|^2(3-\cos{\theta}))}\approx \frac{2g_{TT} g_{AA}}{g_{TT}^2+g_{AA}^2},
\end{equation}

One can see that for similar size couplings, the asymmetry factor can be as large as $\mathcal{O}(1)$.

{\flushleft{\textbf{$C_\chi$ and $P$ conjugation properties of Higher-Dimensional Operators}}}

\begin{table}[h!]
\centering
\renewcommand{\arraystretch}{1.7}
\resizebox{0.9\textwidth}{!}{
\setlength{\tabcolsep}{3pt}
\begin{tabular}{|c|c|c|c|c|c|c|c|c|c|c|c|}
\hline
$C\chi$ / $C\chi P$ 
& $\bar{n}n$ 
& $i\bar{n}\gamma^5 n$ 
& $\bar{n}\gamma_\mu n$ 
& $\bar{n}\gamma_\mu \gamma^5 n$ 
& $\bar{n} \sigma_{\mu\nu} n$ 
& $i \bar{n} \overset{\leftrightarrow}{\partial_\mu} n$ 
& $\bar{n} \overset{\leftrightarrow}{\partial}_\mu\gamma^5 n$ 
& $i \bar{n} \overset{\leftrightarrow}{\partial}_{(\mu} \gamma_{\nu)} n$ 
& $ \bar{n} \overset{\leftrightarrow}{\partial}_{(\mu} \gamma_{\nu)} \gamma^5 n$ \\
\hline
$\bar{\chi} \chi$ 
& $+/+$ & \textcolor{red}{$+/-$} &  &  &  &  &  &  &  \\
\hline
$i \bar{\chi} \gamma^5 \chi$ 
& \textcolor{red}{$+/-$} & $+/+$ &  &  &  &  &  &  &  \\
\hline
$\bar{\chi} \gamma^\mu \chi$ 
&  &  & \textcolor{brown}{$-/-$} & \textcolor{blue}{$-/+$} &  & \textcolor{brown}{$-/-$} & \textcolor{blue}{$-/+$} &  &  \\
\hline
$\bar{\chi} \gamma^\mu \gamma^5 \chi$ 
&  &  & \textcolor{red}{$+/-$} & $+/+$ &  & \textcolor{red}{$+/-$} & $+/+$ &  &  \\
\hline
$\bar{\chi} \sigma^{\mu\nu} \chi$ 
&  &  &  &  & \textcolor{brown}{$-/-$} &  &  & \textcolor{brown}{$-/- $} & \textcolor{blue}{$-/+ $} \\
\hline
$i \bar{\chi} \overset{\leftrightarrow}{\partial}\,^\mu \chi$ 
&  &  & \textcolor{brown}{$-/-$} & \textcolor{blue}{$-/+ $} &  & \textcolor{brown}{$-/-$} & \textcolor{blue}{$-/+ $} &  &  \\
\hline
$\bar{\chi} \overset{\leftrightarrow}{\partial}\,^\mu \gamma^5 \chi$ 
&  &  & \textcolor{blue}{$-/+$} & \textcolor{brown}{$-/-$} &  & \textcolor{blue}{$-/+$} & \textcolor{brown}{$-/-$} &  &  \\
\hline
$i \bar{\chi} \overset{\leftrightarrow}{\partial} \,^{(\mu}\gamma^{\nu )}\chi$ 
&  &  &  &  & $+/+$ &  &  & $+/+$ & \textcolor{red}{$+/-$} \\
\hline
$ \bar{\chi} \overset{\leftrightarrow}{\partial} \,^{(\mu}\gamma^{\nu )}
\gamma^5\chi$ 
&  &  &  &  & \textcolor{blue}{$-/+$} &  &  & \textcolor{blue}{$-/+$} & \textcolor{brown}{$-/-$} \\
\hline
$\chi^\dagger_s \chi_s$ 
& $+/+$ & \textcolor{red}{$+/-$} &  &  &  &  &  &  &  \\
\hline
$i\chi^\dagger_s \overset{\leftrightarrow}{\partial}\,^\mu \chi_s$ 
&  &  & \textcolor{brown}{$-/-$} & \textcolor{blue}{$-/+ $} &  & \textcolor{brown}{$-/-$} & \textcolor{blue}{$-/+ $} &  &  \\
\hline
$\chi^\dagger_s \overset{\leftrightarrow}{\partial}\,^{\mu} \overset{\leftrightarrow}{\partial}\,^\nu\chi_s$ 
&  &  &  &  &  &  &  & \textcolor{black}{$+/+$} & \textcolor{red}{$+/-$} \\
\hline
\end{tabular}}
\caption{$C_\chi$- and $C_\chi P_{\chi+n}$-parities for 4-particle bilinear interaction operator $\mathcal{O}^\chi_{(\mu)} \mathcal{O}^{n(\mu)}$ $\leq$ with up to bilinear order of dim-4. [\textcolor{black}{black} + \textcolor{brown}{brown}] or [\textcolor{blue}{blue} + \textcolor{red}{red}] operator pairs can lead to asymmetric capture. Here $\overset{\leftrightarrow}{\partial}_\mu \equiv \overset{\rightarrow}{\partial}_\mu - \overset{\leftarrow}{\partial}_\mu$ denotes derivative difference between right and left derivatives, and $(\mu,\nu)$ means either fully-symmetric or anti-symmetric permutation of Lorentz indices. The $\chi_s$ terms are scalar DM bilinears.  \label{tab:operators-full}}
\end{table}

From an effective field theory perspective, higher-dimensional operators are also possible, which can be constructed from the Dirac gamma matrices and possible derivatives acting on fermionic and bosonic fields. We include dim-4 operators, with at most one-derivative for fermion bilinears, and two-derivative for scalar bilinears. 
Here we define $\overset{\leftrightarrow}{\partial}_\mu \equiv \overset{\rightarrow}{\partial}_\mu - \overset{\leftarrow}{\partial}_\mu$ as the difference between the left and right derivatives, which is useful for writing down self-hermitian bilinear operators. Meanwhile we use $(\mu,\nu)$ to represent either fully-symmetric permutation of Lorentz indices (for the energy-stress tensor-like operators), or fully-asymmetric permutation. We use $\chi_s(\chi_s^\dagger)$ to denote the scalar DM field (and its complex conjugate), and $\chi(\bar{\chi})$ for the Dirac fermionic DM field (and its Hermitian conjugate). 

As defined in the main text, $C_\chi$ is the charge conjugation parity of the DM bilinear operator, and $C_\chi P_{\chi+n}$ is with an additional parity transformation on both DM and nucleon operators. We list the results in Table.\ref{tab:operators-full}, and [\textcolor{black}{black} + \textcolor{brown}{brown}] or [\textcolor{blue}{blue} + \textcolor{red}{red}] operator pairs can have different $C_\chi$ and $C_\chi P_{\chi+n}$ parities, which can generate the asymmetry between DM and anti-DM scattering cross-sections.

\section{Capture, Thermalization, Collapse of Dark Matter into a Black Hole and its subsequent evolution}

To examine the typical timescales required for DM capture and settling to the NS center, thermalization, and eventual collapse, we first specify the parameter region of interest. We consider a typical old NS with mass $1.5M_\odot$, radius $12$km and age $10^{10}$yr. For dark matter particles, the two key parameters are the DM mass $m_\chi$ and the DM-nucleon scattering cross-section $\sigma_{\chi n}$. We focus specifically on heavy DM masses within the range $m_\chi \in [10^5,10^{12}]\,{\rm GeV}$, and weak cross-sections in the range $\sigma_{\chi n} \in [10^{-48},10^{-40}]\,{\rm cm}^2$. Stronger cross-sections can be trivially addressed by assuming a maximal capture probability $p_{\rm cap}=100\%$ and negligible thermalization time.

\par Considering the symmetric DM scenario, in which the DM and anti-DM particles have identical mass $m_\chi$ and each constitutes equal fraction of the local dark matter halo density ($n_\chi = n_{\tilde{\chi}} = n_{\rm DM}/2$, assuming that DM is saturated by $\chi$ and $\tilde \chi$), the evolution equations for the total number of each species inside a star are:
\begin{align}
\frac{\D N_\chi}{\D t} &= C_\chi - N_\chi\frac{N_{\tilde{\chi}}}{V_{\rm th}}\langle \sigma_{\chi\tilde{\chi}} v\rangle, \\[5pt]
\frac{\D N_{\tilde{\chi}}}{\D t} &= C_{\tilde{\chi}} - N_{\tilde{\chi}}\frac{N_{\chi}}{V_{\rm th}}\langle \sigma_{\chi\tilde{\chi}} v\rangle,
\end{align}
where $V_{\rm th}$ is the volume of the thermalized DM sphere residing inside the NS~\cite{Goldman:1989nd,Page:2006ud,Bertoni:2013bsa,Bramante:2023djs}, and $\langle \sigma_{\chi\tilde{\chi}} v\rangle$ is the thermally averaged annihilation cross-section, assumed to be effective at sufficiently high (anti-)DM densities. Under typical conditions, such annihilation processes prevent substantial build-up of either DM or anti-DM populations within the NS. However, a non-zero asymmetry parameter $\mathcal{A}$ induces a faster capture rate for one species over the other, leading to a net accumulation of DM particles even after annihilation. By subtracting the above equations, we obtain the evolution for the residual DM particle number:
\begin{equation}
   \frac{\D N^{\rm res}}{\D t} = \frac{\D (N_{\chi}-N_{\tilde{\chi}})}{\D t} = C_\chi - C_{\tilde{\chi}} \overset{p_\chi\rightarrow 0}{\approx} \frac{\mathcal{A}}{1+\mathcal{A}} n_{\rm DM} v_{\rm rel} \pi b^2 p_{\rm cap},
\end{equation}
noting that $C_{\chi/\tilde{\chi}}$ is defined in the main text.

Once the number of DM particles accumulated in the center of a NS in form of a thermalized spherical population reaches a critical threshold, an irreversible gravitational collapse into a mini BH occurs. The criterion for this collapse can be characterized by three simultaneous conditions: (1) the thermal gas pressure response is insufficiently rapid to counteract gravitational infall, known as Jeans instability; (2) no stable gravitationally bound orbit exists for the thermalized DM particles under gravity alone; (3) the quantum degeneracy pressure of DM particles cannot support their self-gravity, defined by the Chandrasekhar limit. Among these three conditions, the Chandrasekhar limit imposes the most stringent requirement~\cite{Page:2006ud,Narain:2006kx,Acevedo:2020gro,Bramante:2023djs,Bhattacharya:2023stq,Ge:2024prt}, thus setting the critical mass for gravitational collapse as:
\begin{equation}
    M_{\rm crit} = \omega\,\frac{M_{\rm pl}^3}{m_\chi^2} 
    = 6.2\times 10^{-17}M_\odot\left(\frac{\omega}{0.384}\right)\left(\frac{10^8\,{\rm GeV}}{m_\chi}\right)^{2},
\end{equation}
where $\omega$ is a dimensionless factor from the integration of the TOV equation~\cite{Oppenheimer:1939ne}. The original Chandrasekhar limit gives $\omega\approx 3$ for degenerate electron gas in white dwarfs~\cite{Chandrasekhar:1931ih}. Most existing literature in the field uses $\omega = 1$ as an approximation~\cite{Page:2006ud,Acevedo:2020gro,Bramante:2023djs,Bhattacharya:2023stq,Ge:2024prt}. Here we take $\omega = 0.384$ from ~\cite{Narain:2006kx}, acquired from a full GR integration of TOV equations for heavy fermionic degenerate gas with no self-interaction. The corresponding timescale required for the NS to accrete sufficient DM to reach this critical mass is then straightforwardly calculated as:
\begin{equation}
    t_{\rm acc} = \frac{M_{\rm crit}/m_\chi}{\D N^{\rm res}/\D t}.
\end{equation}

The thermalization process for DM particles can be divided into two distinct stages. In the initial stage, when a DM particle first becomes gravitationally bound, most of its orbit remains outside the NS. During this phase, the particle loses energy only inside the NS. As the DM particle gradually cools, its orbit continuously shrinks until eventually it is fully contained within the NS, marking the transition into the second stage. As shown explicitly in~\cite{Bertoni:2013bsa,Acevedo:2020gro}, most of the total thermalization time is spent during the last few scattering events, after the particle has already settled deep within the NS core. Thus, the thermalization timescale can be well approximated as~\cite{Bertoni:2013bsa}:
\begin{equation}
\begin{aligned}
    t_{\rm th}&\approx \frac{105\pi^2}{16\,m_n\,\sigma_{n\chi}\,T_*^2}\frac{m_\chi/m_n}{(1+m_\chi/m_n)^2}\\[5pt]
    &\approx 1.88\times 10^{7}\,{\rm yr}\left(\frac{10^{-45}\,{\rm cm}^2}{\sigma_{\chi n}}\right)\left(\frac{1000\,{\rm K}}{T_*}\right)^2 \frac{m_\chi/m_n}{(1+m_\chi/m_n)^2}.
\end{aligned}
\end{equation}
Given that $m_n\ll m_\chi$ in our parameter region of interest, the thermalization time is negligible compared to the typical lifetime of NSs.

For very large DM–nucleon cross-sections, frequent collisions between DM particles and nucleons could potentially introduce a significant viscous drag force, slowing the DM particle's drift toward the NS core. However, such viscous drag only becomes relevant for cross-sections substantially greater than $10^{-20}\,{\rm cm}^2$ and thus is entirely irrelevant for our chosen parameter range~\cite{Acevedo:2020gro}. Once a sufficient number of DM particles have been accumulated and thermalized within the thermal sphere, the subsequent gravitational collapse occurs rapidly, on timescales determined by specific cooling mechanisms and radiative channels. Generally, this collapse timescale is negligible in comparison to the age of the NS, $t_{\rm age}$~\cite{Bramante:2015cua}.

After the DM collapse, a mini BH with initial mass $M_{\rm crit}$ will form, and it starts to accrete the NS matter and evaporate simultaneously. Some recent papers also considered the accretion of constant incoming DM flux to the central mini BH ~\cite{Garani:2017jcj,Acevedo:2020gro,Ray:2023auh,Bramante:2023djs,Ge:2024prt}. The evolution of the mass of the mini BH can be described as:
\begin{equation}
\frac{\mathrm{d} M_{\mathrm{BH}}}{\mathrm{d} t}=\frac{4 \pi \rho_{*}\left(G M_{\mathrm{BH}}\right)^2}{c_{\rm s}^3}+m_\chi\frac{\D N^{\text{res}}}{\D t}-\frac{f\left(M_{\mathrm{BH}}\right)}{\left(G M_{\mathrm{BH}}\right)^2},
\end{equation}
with $\rho_*$ the core density of NS and $c_s$ the core sound speed. Here the first term describes the Bondi Accretion of NS matter, the second term the DM accretion and the third term is the Hawking Radiation effect, with $f(M_{\rm BH})$ the Page factor~\cite{MacGibbon:1991tj}.  Thus there exists a critical BH mass for accretion-evaporation balance. Note that the DM accretion rate is much smaller than the Bondi accretion rate here, giving the following estimate for the runaway condition:
\begin{equation}
    M_0\approx \frac{1}{G}\sqrt[4]{\frac{f(M_0)c_*^3}{4\pi \rho_*}} \approx 4.2\times 10^{-20}M_\odot\left(\frac{f(M_0)}{1/(74\pi)}\right)^{1/4}\left(\frac{c_*}{0.5c}\right)^{3/4}\left(\frac{10^{15}{\rm g}/{\rm cm}^3}{\rho_*}\right)^{1/4}.
\end{equation}
Here $f(M_{\rm BH})=1/(74\pi)$ if all the Standard Model states are considered to be emitted ~\cite{MacGibbon:1991tj,Ray:2023auh}. Such critical mass would become 
$2.1\times 10^{-20}M_\odot$ for the Page factor $f(M_0)=1/(1135\pi)$, when only photon and neutrino radiation is accounted for the emission. Black holes created from dark matter collapse below the critical mass would eventually Hawking-radiate away, terminated the development of a BH and preserving the NS. This black hole mass scale corresponds to a Hawking radiation temperature $T_{\rm BH} = (8\pi G M_{\rm BH})^{-1} $ of $\approx 127\sim 253{\rm MeV}$, thus only some of the standard model particles can be produced in the Hawking radiation process, rendering a Page factor in between these two benchmark values. Here we include both values for our calculations and results in the main text, differing from existing literature~\cite{deLavallaz:2010wp,Bramante:2017xlb} which take the most aggressive assumption on the Page factor $f(M_0) =1/(15460\pi)$ considering only photon emissions.

Above the critical BH mass, the Hawking evaporation effect can be neglected, and the growth time of the mini BH can be estimated as the Bondi accretion process:
\begin{equation}
    \frac{\D M_{\rm BH}}{\D t} = \frac{4\pi \rho_* G^2 M_{\rm BH}^2}{c_*^3}.
\end{equation}
Such process has a definite timescale of growth, as the explicit solution to this equation is given by
\begin{equation}
    M_{\rm BH}(t) = \frac{M_{\rm crit}}{1-M_{\rm crit}4\pi \rho_* G^2 t/{c_*^3}},
\end{equation}
where the growth time scale is:
\begin{equation}
\begin{aligned}
    t_{\rm growth}= \frac{c_*^3}{4\pi \rho_* G^2 M_{\rm crit}}
    \approx 1.01\times 10^{8}{\rm yr}\left(\frac{10^{-20}M_\odot}{M_{\rm crit}}\right)\left(\frac{c_*}{0.5c}\right)^{3}\left(\frac{10^{15}{\rm g}/{\rm cm}^3}{\rho_*}\right).
\end{aligned}
\end{equation}
Note that when the BH gets too heavy, the radiation produced in the accretion process may become strong enough to heat up the fluid and provide extra pressure support for the in-falling stellar matter, which can potentially slow the accretion rate to the Eddington-limited rate. However such effect only happens when the BH is rather  large, and would not affect the total growth time dominated by initial stages of accretion. Thus, such an effect can be safely ignored ~\cite{Bellinger:2023wou,Ge:2024prt} for the purpose of deriving NS$\to$BH bounds.

\end{document}